\documentclass[conference]{IEEEtran}
\IEEEoverridecommandlockouts
\usepackage{cite}
\usepackage{amsmath,amssymb,amsfonts}
\usepackage{algorithmic}
\usepackage{graphicx}
\usepackage{textcomp}
\usepackage{xcolor}

\usepackage{soul}
\setstcolor{red}

\newcounter{tempcounter_1}
\newcounter{tempcounter_2}
\makeatletter
\newcommand{\linebreakand}{%
\end{@IEEEauthorhalign}
\hfill\mbox{}\par
\mbox{}\hfill\begin{@IEEEauthorhalign}
}
\makeatother

\def\BibTeX{{\rm B\kern-.05em{\sc i\kern-.025em b}\kern-.08em
    T\kern-.1667em\lower.7ex\hbox{E}\kern-.125emX}}
\begin{document}

\title{Preventing output saturation in active noise control: An output-constrained Kalman filter approach}

\author{
\IEEEauthorblockN{Junwei Ji\textsuperscript{1}}
\IEEEauthorblockA{
\textit{Nanyang Technological University}\\
Singapore \\
junwei002@e.ntu.edu.sg}
\and
\IEEEauthorblockN{Dongyuan Shi\textsuperscript{2}}
\IEEEauthorblockA{
\textit{Nanyang Technological University}\\
Singapore \\
dshi003@e.ntu.edu.sg}
\and
\IEEEauthorblockN{Boxiang Wang\textsuperscript{3}}
\IEEEauthorblockA{
\textit{Nanyang Technological University}\\
Singapore \\
boxiang001@e.ntu.edu.sg}
\linebreakand
\IEEEauthorblockN{Xiaoyi Shen\textsuperscript{4}}
\IEEEauthorblockA{
\textit{Nanyang Technological University}\\
Singapore \\
xiaoyi.shen@ntu.edu.sg}
\and
\IEEEauthorblockN{Zhengding Luo\textsuperscript{5}}
\IEEEauthorblockA{
\textit{Nanyang Technological University}\\
Singapore \\
luoz0021@e.ntu.edu.sg}
\and
\IEEEauthorblockN{Woon-Seng Gan\textsuperscript{6}}
\IEEEauthorblockA{
\textit{Nanyang Technological University}\\
Singapore \\
ewsgan@ntu.edu.sg}
}

\maketitle

\begin{abstract}
The Kalman filter (KF)-based active noise control (ANC) system demonstrates superior tracking and faster convergence compared to the least mean square (LMS) method, particularly in dynamic noise cancellation scenarios. However, in environments with extremely high noise levels, the power of the control signal can exceed the system's rated output power due to hardware limitations, leading to output saturation and subsequent non-linearity. To mitigate this issue, a modified KF with an output constraint is proposed. In this approach, the disturbance treated as an measurement is re-scaled by a constraint factor, which is determined by the system's rated power, the secondary path gain, and the disturbance power. As a result, the output power of the system, i.e. the control signal, is indirectly constrained within the maximum output of the system, ensuring stability. Simulation results indicate that the proposed algorithm not only achieves rapid suppression of dynamic noise but also effectively prevents non-linearity due to output saturation, highlighting its practical significance.
\end{abstract}


\section{Introduction}
\label{sec:intro}
In recent years, more attention has been brought to the issue of noise pollution, which not only affects the quality of life but also causes many health problems. Active noise control (ANC) systems, which employ secondary sources to generate the anti-noise that diminishes the undesirable sound \cite{KuoANC1999,Elliott1993ANC,Kajikawa2012recent}, have become increasingly essential in various applications, particularly in environments where traditional passive noise control methods are insufficient \cite{lam2021ten}, such as headrest \cite{shen2023implementations,hasegawa2017headrest,buck2018performance,zhang2022robust}, vehicles \cite{jung2019local,jia2020hybrid,jiang2021modified,wang2023experimental}, open aperture windows \cite{lee2021review,huang2011active,murao2017mixed,shi2019practical}, etc. Among the various algorithms used for ANC \cite{Morgan1980FxLMS,lee2009subband,luo2023delayless,zhang2023deep}, the Kalman filter (KF) has gained considerable attention due to its superior tracking capabilities and faster convergence rates, especially in scenarios involving dynamic noise. 

Despite the advantages of the KF-based ANC system, it also faces significant challenges in environments with extremely high noise levels. Due to the inherent hardware limitations, the output power of the signal should be constrained to the rated output power of the system. Otherwise, the large output power may overdrive the secondary source, leading to the output saturation issue and subsequent non-linearity. Clipping the signal is a simple method but exacerbates the nonlinearity of the system, causing the control filter to overrun \cite{qiu2001study}. Conventional filtered reference least mean square (FxLMS) based algorithms are developed to avoid divergence \cite{guo2024survey}, such as re-scaling the output signal \cite{qiu2001study} or modifying the update direction of the control filter \cite{shi2019two}. These algorithms on the basis of output amplitude constraints may compromise error signal quality and introduce amplitude distortion. To mitigate these issues, a strategy that restricts the output power is proposed. Leaky FxLMS \cite{tobias2005leaky} and minimum output variance FxLMS (MOVFxLMS) \cite{shi2021optimal} introduce a penalty factor confining the update of the control filter, where the penalty factor can be determined by the rated output power of the system \cite{shi2019optimal,lai2023mov}. These FxLMS-based algorithms basically restrain the output power of the system indirectly by changing the gradient of the control filter updating. However, these methods do not apply to the KF-based ANC since the KF provides the optimal estimate of the control filter through a linear combination of the previous estimated control filter and the newest observed disturbance \cite{diniz1997adaptive}.

To the best of our knowledge, existing studies have only attempted to apply KF to the ANC system \cite{lopes1999kalman,petersen2008kalman,van2013multi,fraanje2005fast,fabry2018active,luo2023gfanc}, but have rarely considered the problem of output saturation. To tackle this issue, a Kalman filter with output power constraint (KF-OPC) approach is proposed. It should be noted that the disturbance signal is regarded as the measurement to correct the control filter in the KF-based ANC system. Therefore, a constraint factor is introduced to re-scale the disturbance signal at the correct stage, where the constraint factor is determined by the system's rated output power, the secondary path gain and the disturbance power. As a result, the power reduction of the measurement indirectly limits the output power of the system to within the rated output power, making the KF-based ANC system more practical. The simulation results demonstrate that the proposed KF-OPC algorithm not only inherits the advantages of KF but also effectively prevents output saturation and thus avoids the phenomenon of non-linearity.

The remaining sections of this paper are organized as follows: Section \ref{sec:proposed} briefly reviews the KF-based ANC system followed by the power-constrained KF approach. In Section \ref{sec:simulation}, several numerical simulations are conducted to illustrate the proposed method's effectiveness in overcoming the output saturation problem. Finally, conclusions are drawn in Section \ref{sec:conclusion}.

\begin{figure}[!t]
    \centering
    \includegraphics[width = 8.5cm]{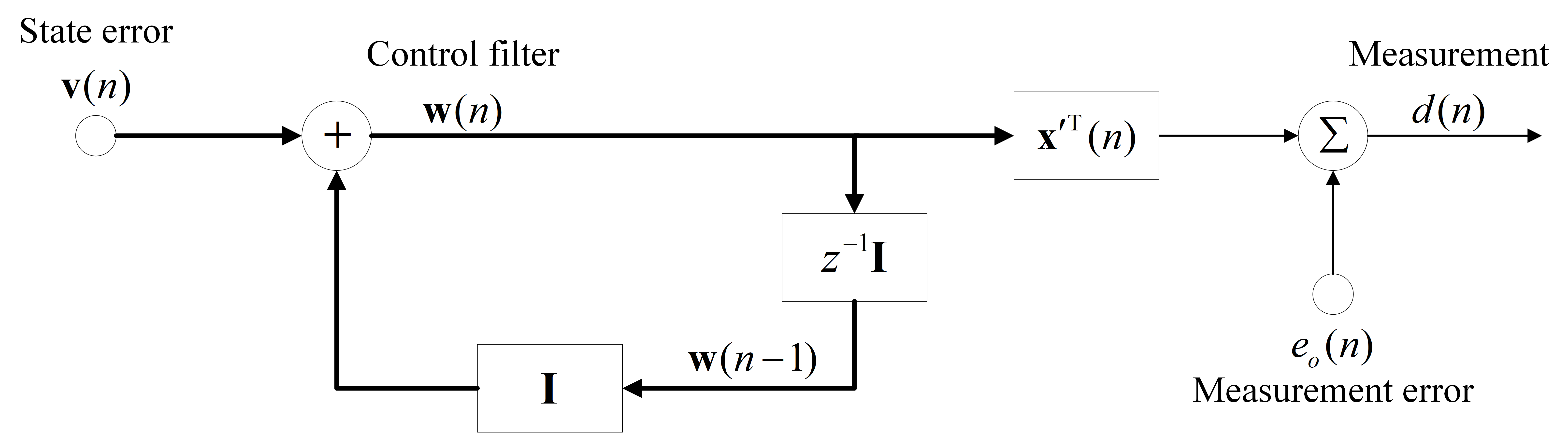}
    \caption{The state space model for the ANC system}
    \label{fig:1 SSMANC}
\end{figure}
\vspace{-0.3cm}
\section{Methodology}
\label{sec:proposed}
\subsection{ANC with Kalman filter (ANC-KF)} \label{ssec:reviewkfanc}


Conventional FxLMS-based ANC system solves the problem of estimating the disturbance signal $d(n)$, while the KF-based ANC system provides the optimal estimate of the control filter $\mathbf{w}(n)$ \cite{diniz1997adaptive}. To apply the KF to the ANC system, a state space model for the ANC system is developed as depicted in Fig.~\ref{fig:1 SSMANC}, where $\mathbf{I}$ represents the identity matrix of $L\times L$ and $L$ is the length of control filter. $\mathbf{x'}(n)$ is the filtered reference vector with the length of $L$, which is obtained by filtering the reference signal $x(n)$, through the estimated secondary path $\hat{s}(n)$.

According to the state space model, the ANC system can be described as
\begin{subequations}\label{eq1:SEOE}
    \begin{align}
        \mathbf{w}(n) = \mathbf{w}(n-1) + \mathbf{v}(n), \label{eq1a:se}\\
        d(n) = \mathbf{x'}^\mathrm{T}(n)\mathbf{w}(n) + e_o(n), \label{eq1b:oe}
    \end{align}
\end{subequations}
where \eqref{eq1a:se} and \eqref{eq1b:oe} denote as state equation and observation equation, respectively. The control filter $\mathbf{w}(n)$ regarded as a state vector is generated from the previous state $\mathbf{w}(n-1)$, and from the state error $\mathbf{v}(n)$. The disturbance signal $d(n)$ treated as observed or measured signal originates from the anti-noise signal $\mathbf{x'}^\mathrm{T}(n)\mathbf{w}(n)$ and measurement error represented as $e_o(n)$.

Once the relevant model description is available, the Kalman filter can thus be derived to seek the optimal estimate of the control filter, denoted as $\mathbf{\hat{w}}(n|n)$. In general, the Kalman filter provides an optimal estimate of the system state through two processes: time update (or predict) and measurement update (or correct). 

In the time update procedure, the current state vector is predicted by the previous state as
\begin{equation}\label{eq2:predictstate}
    \mathbf{\hat{w}}(n|n-1) = \mathbf{\hat{w}}(n-1|n-1),
\end{equation}
where $\mathbf{\hat{w}}(n|n-1)$ is a prior state estimate, meaning that the estimate of $\mathbf{w}(n)$ is obtained when the current measurement is not available. $\mathbf{\hat{w}}(n-1|n-1)$ is the previous optimal estimate, which can also be considered as a posterior state estimate. Then, the covariance of the state-prediction error is defined as
\begin{equation}\label{eq3:predicterror}
    \mathbf{P}(n|n-1) = \mathbb{E}\{[\mathbf{w}(n)-\mathbf{\hat{w}}(n|n-1)][\mathbf{w}(n)-\mathbf{\hat{w}}(n|n-1)]^{\mathrm{T}}\}.
\end{equation}
Based on \eqref{eq1a:se} and \eqref{eq2:predictstate}, the covariance of the state-prediction error can be expressed in an iteration formula as
\begin{equation}\label{eq4:predicterroriter}
    \mathbf{P}(n|n-1) = \mathbf{P}(n-1|n-1) + \mathbf{r}(n)\mathbf{I},
\end{equation}
where $\mathbf{r}(n)$ denotes the variance of the state error $\mathbf{v}(n)$.

If the newest measurement $d(n)$ is available, the next step is to correct the prior state estimate $\mathbf{\hat{w}}(n|n-1)$. Therefore, the optimal estimate of the control filter can be updated as
\begin{equation}\label{eq5:update}
    \mathbf{\hat{w}}(n|n) = \mathbf{\hat{w}}(n|n-1) + \mathbf{K}(n)[d(n) - \mathbf{x'}^\mathrm{T}(n)\mathbf{\hat{w}}(n|n-1)],
\end{equation}
where $\mathbf{K}(n)$ represents the Kalman gain matrix which characterizes the proportion of model prediction error to measurement error in the measurement update process. A larger Kalman gain indicates a greater confidence in the measurements. On the other hand, the Kalman gain is also related to the covariance of the optimal state estimate error $\mathbf{P}(n|n)$. To minimize $\mathbf{P}(n|n)$, the Kalman gain matrix under optimal estimation conditions can be calculated as
\begin{equation}\label{eq6:kalmangain}
    \mathbf{K}(n) = \mathbf{P}(n|n-1)\mathbf{x'}(n)[\mathbf{x'}^\mathrm{T}(n)\mathbf{P}(n|n-1)\mathbf{x'}(n) + q(n)]^{-1},
\end{equation}
where the $q(n)$ is the variance of the measurement error, defined as
\begin{equation}\label{eq7:measureerror}
    q(n) = \mathbb{E}[e_o^2(n)],
\end{equation}
where $\mathbb{E}[\cdot]$ represents the expectation operation. Finally, the covariance of the optimal state estimate error $\mathbf{P}(n|n)$ is updated as
\begin{equation}\label{eq8:stateerrorupdate}
    \mathbf{P}(n|n) = [\mathbf{I} - \mathbf{K}(n)\mathbf{x'}^\mathrm{T}(n)]\mathbf{P}(n|n-1).
\end{equation}

\begin{figure}[!t]
    \centering
    \includegraphics[width = 8.5cm]{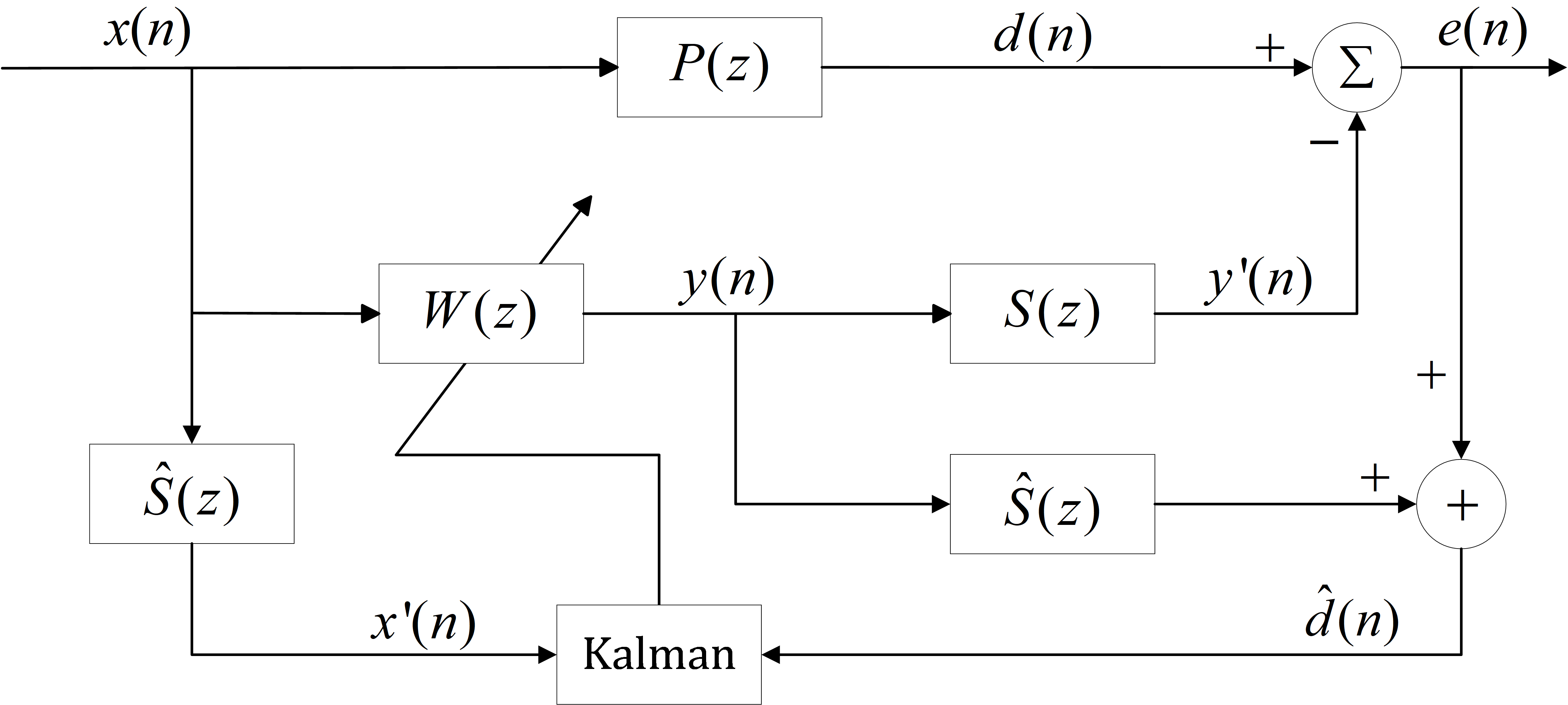}
    \caption{The block diagram of ANC using KF approach, where $P(z)$ and $S(z)$ are the primary and secondary path, respectively. $\hat{S}(z)$ denotes the estimated secondary path and $W(z)$ is the control filter. $x(n)$, $y(n)$, $y'(n)$, and $e(n)$ are the reference signal, control signal, anti-noise, and residual error.}
    \label{fig:3 KFANC}
\end{figure}

Therefore, \eqref{eq2:predictstate},\eqref{eq4:predicterroriter},\eqref{eq6:kalmangain},\eqref{eq5:update} and \eqref{eq8:stateerrorupdate} are the five main equations for the ANC-KF \cite{yu2024implementation}. In the practical application of ANC, the modified ANC structure is used to realize the ANC-KF as shown in Fig.~\ref{fig:3 KFANC}, since the disturbance signal as a measurement cannot be obtained directly. The output of the Kalman filter $\mathbf{\hat{w}}(n|n)$ is then used directly as a true control filter $\mathbf{w}(n)$ to generate the control signal.

\subsection{Kalman filter with output power constraint for ANC system}\label{ssec:KF-OPC}

The Kalman filter provides the optimal estimation of the control filter and hence it has a fast convergence speed. However, in an environment of large amplitude disturbance signal, the hardware capability of the ANC system is confined to not perfectly output the high-power control signal, which can easily cause output saturation. The distortion of the output signal leads to the distortion of the measurement, which further affects the update of the KF. Therefore, we propose a KF with output power constraint (KF-OPC) method to address such a problem, making the KF-based ANC system more practical.

In contrast to FxLMS-based ANC systems where the update gradient can be changed to suppress the power of the output signal, KF-ANC updates the controller with measurement. Therefore, it is difficult to apply the traditional power-constrained FxLMS algorithm directly to the KF-ANC. It is worth noting that because the KF-ANC corrects the control filter by means of measurement, we can re-scale the measurement before feeding it into the KF. By modifying the amplitude of the measurements, the power of the measurement is also altered, thus confining the update of the KF, and the power of the output signal as a result can be constrained. Therefore, we introduce a constraint factor $\alpha$ into \eqref{eq5:update} resulting in
\begin{equation}\label{eq9:updateconstraint}
    \mathbf{\hat{w}}(n|n) = \mathbf{\hat{w}}(n|n-1) + \mathbf{K}(n)[\alpha \hat{d}(n) - \mathbf{x'}^\mathrm{T}(n)\mathbf{\hat{w}}(n|n-1)],
\end{equation}
where the disturbance signal $d(n)$ is replace with the estimated disturbance signal $\hat{d}(n)$ in real application.

\begin{figure}[!t]
    \centering
    \includegraphics[width = 8.5cm]{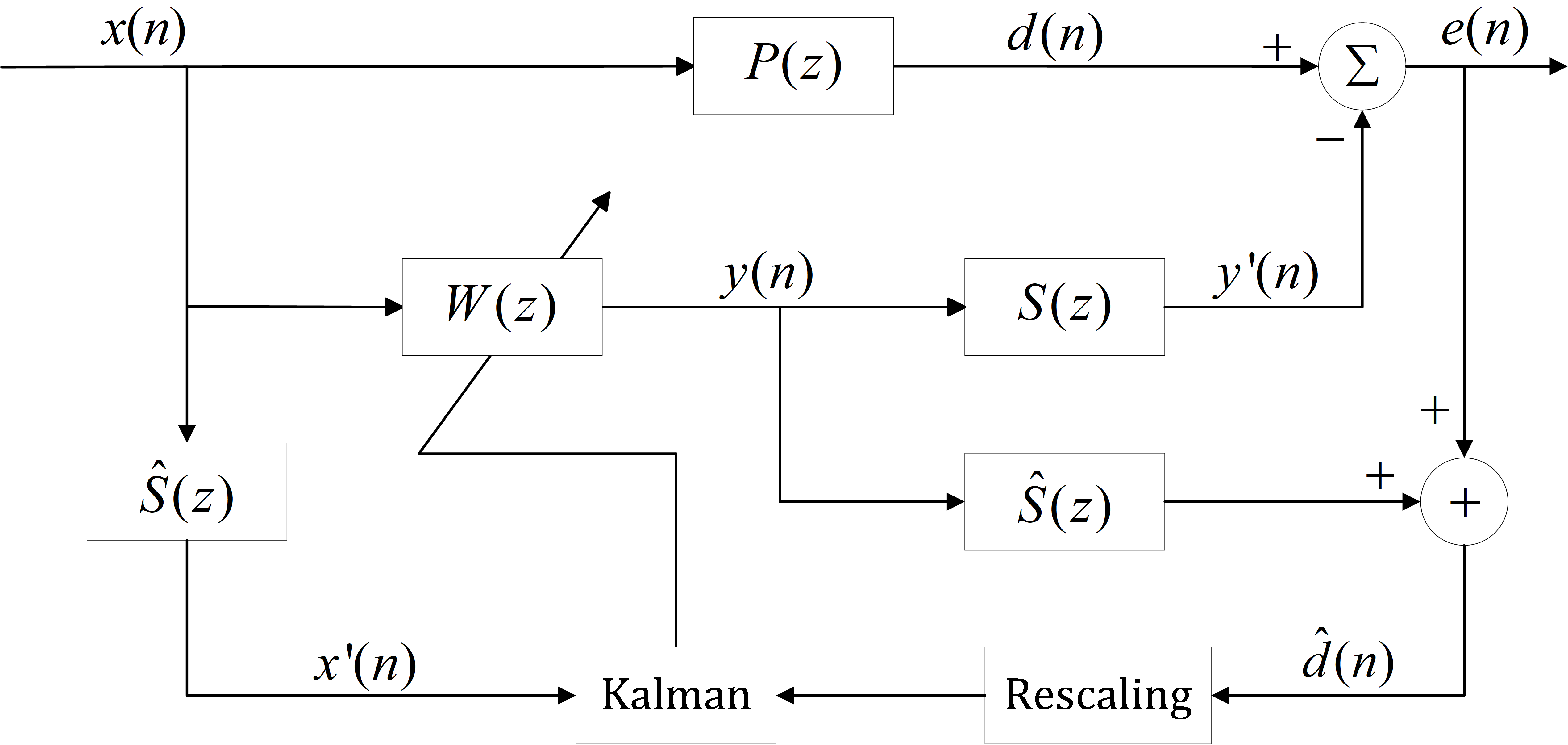}
    \caption{The block diagram of KF with output power constraint (KF-OPC) method for the ANC system to avoid output saturation problem}
    \label{fig:4 KFOPC}
\end{figure}

In order to set a reasonable value for $\alpha$, we exploit the relationship between the rated output power of the system and the actual power of the disturbance signal to determine it. First, we define the rated output power of the system as $\rho_o$ and the power of the disturbance signal as
\begin{equation}\label{eq10:Disturbancepower}
    \delta^2_d = \mathbb{E}[d^2(n)].
\end{equation}
In the ANC system, the output signal needs to be passed through the secondary path to generate an anti-noise signal in order to cancel the disturbance signal in the desired area. The secondary path will provide gain to the output signal, so we assume that the gain of the secondary path is $Gs$. According to \cite{shi2021optimal}, the secondary path gain is related to the frequency distribution of the signal. In order to estimate $Gs$, we derive it using the ratio of the filtered reference signal to the reference signal,
\begin{equation}\label{eq11:Gs}
    Gs  = \frac{\mathbb{E}[x'^{2}(n)]}{\mathbb{E}[x^{2}(n)]}.
\end{equation}
Theoretically, the anti-noise signal power obtained from the maximum output power of the system through the secondary path should match the power of the disturbance signal that is being constrained. For this purpose, the following relationship can be obtained:
\begin{equation}\label{eq12:relation}
    \rho_o Gs = \delta^2_d\alpha^2.
\end{equation}
Therefore, the constraint factor can be chosen as
\begin{equation}\label{eq13:alpha}
    \alpha = \min\{ \sqrt{\frac{\rho_oGs}{\delta^2_d}},1\}.
\end{equation}
When the output power of the system is sufficient to cancel all noise signals, we do not constrain the KF measurement and set $\alpha$ to 1.

Therefore, the block diagram of the proposed KF-OPC method is illustrated in Fig.~\ref{fig:4 KFOPC}. As in conventional KF-ANC, we first exploit the structure of the modified ANC to obtain the measurement, i.e., the estimated disturbance signals. For the purpose of preventing output saturation in a loud noise environment, the optimal control filter is re-estimated by re-scaling the measurement, which indirectly constrains the power of the output signal and also ensures the stability of the system.

\vspace{-0.1cm}
\section{Numerical Simulation}
\label{sec:simulation}

\subsection{Performance on overcoming output saturation}
To evaluate the efficacy of the proposed KF-OPC algorithm in mitigating distortions caused by the amplifier's output saturation, a simplified amplifier model \cite{shi2019optimal} is used after the control filter. The primary and secondary paths are band-pass filters from 20 to 5,000Hz with length of 128 and 32, respectively. The 400Hz tonal noise serves as the primary noise so that the distortion of the amplifier output is more readily observed. Moreover, the power constraint is chosen as $\rho_0 = 1$ resulting in the amplitude constraint with $\sqrt{2}$. 

\begin{figure}[!t]
    \centering
    \includegraphics[width = 8.5cm,height = 7cm]{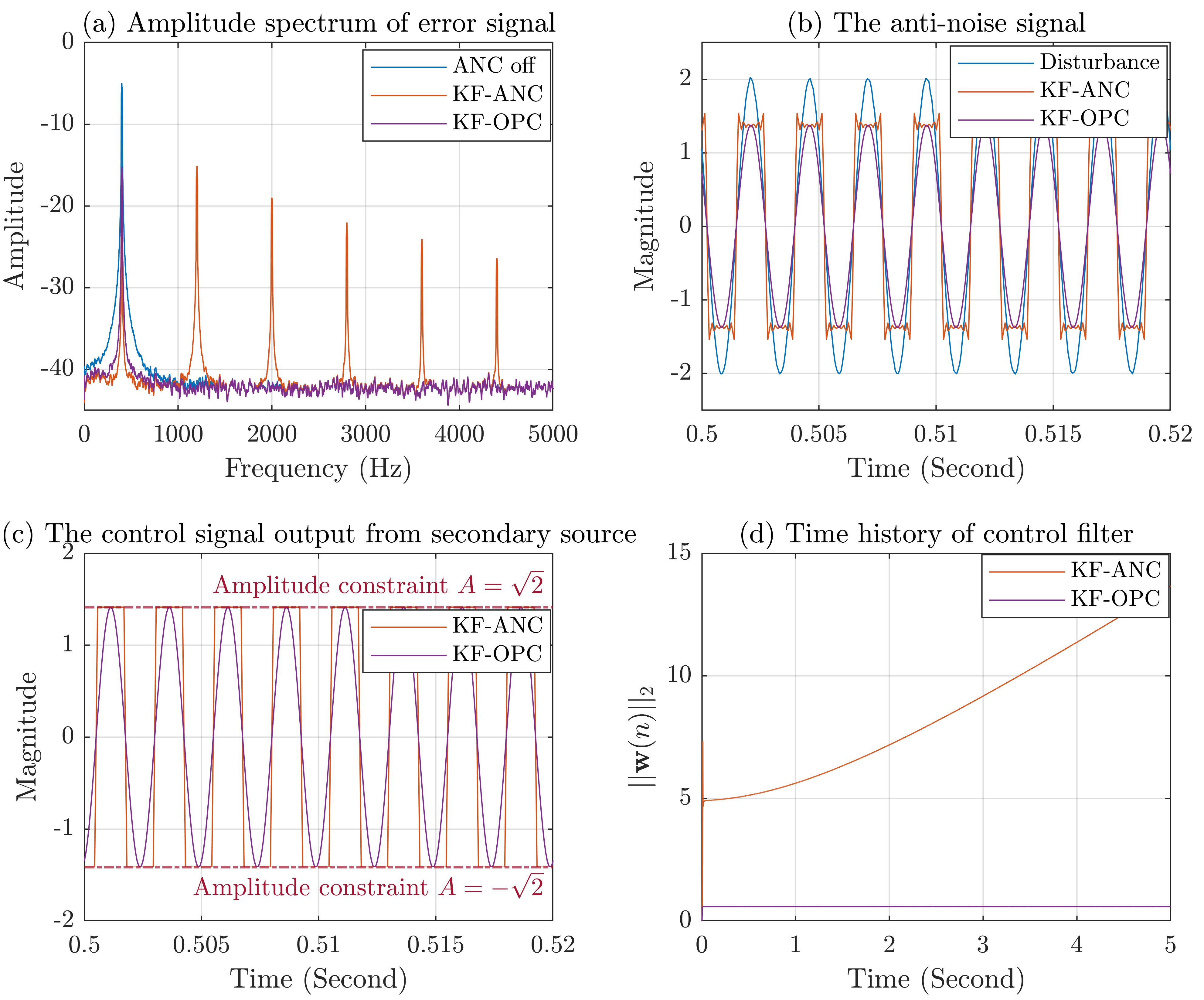}
    \caption{Noise reduction performance with limited output power: (a) Power spectrum of the error signal in different algorithms, (b) the anti-noise signal waves of different algorithms, (c) the control signal waves output from secondary source, and (d) the time history of one control filter weight in different algorithms.}
    \label{fig:5 case1}
\end{figure}

From Fig.~\ref{fig:5 case1}(a), it can be seen that the conventional KF-ANC produces many harmonics leading to its noise reduction performance being inferior to that of the proposed KF-OPC algorithm. Through Fig.~\ref{fig:5 case1}(b) it can be observed that it is because of the significant distortion of the anti-noise signal of the KF-ANC that leads to the generation of harmonics. Further examination of the output of the secondary source in Fig.~\ref{fig:5 case1}(c) reveals that the control signal exceeds the maximum value that can be output by the amplifier resulting in partial truncation of the signal and hence a series of undesirable effects. However, the proposed KF-OPC algorithm avoids the output saturation problem and ensures signal integrity. Figure.~\ref{fig:5 case1}(d) further illustrates that the control filter of KF-ANC overflows due to output saturation, resulting in the divergence of the adaptive systems.

\subsection{Performance on broadband noise}
\begin{figure}[!t]
    \centering
    \includegraphics[width = 8.5cm,height=7cm]{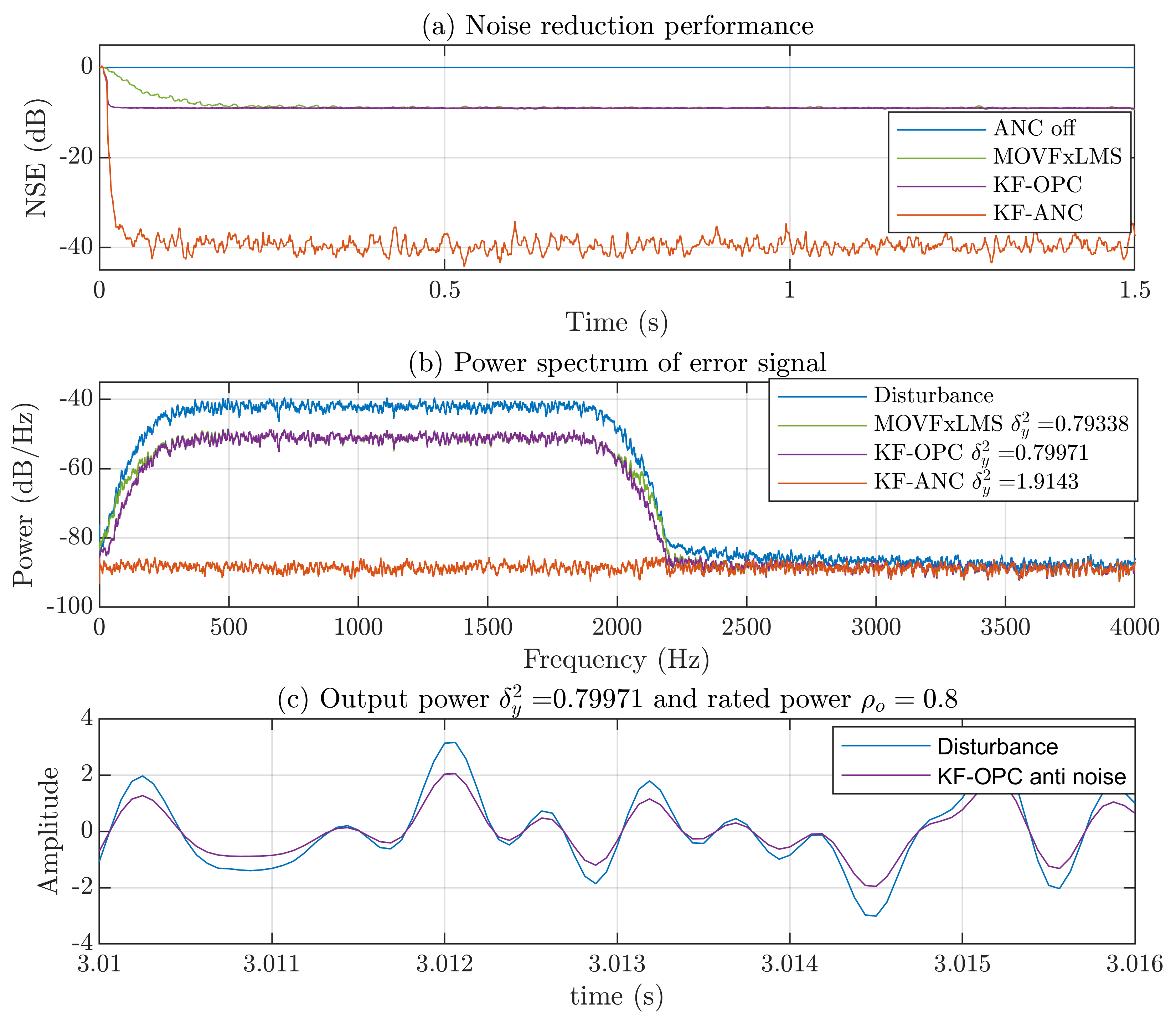}
    \caption{Noise reduction performance with broadband noise: (a) Noise reduction performance with different algorithms, (b) The power spectrum of error signal with different algorithms, and (c) The disturbance and KF-OPC anti-noise waveform with the output power $0.79971$ and the rated power $\rho_o = 0.8$.}
    \label{fig:6 case2}
\end{figure}

In this case, the primary noise changes to the broadband noise ranging from 200 to 2,000Hz, and the system's rated power $\rho_o$ is set to 0.8. The normalized squared error (NSE) \cite{li2023distributed} shown in Fig.~\ref{fig:6 case2} (a) illustrates that the proposed KF-OPC algorithm with the introduced constraint factor does not affect the overall convergence speed, and is faster than MOVFxLMS for noise reduction. Although the KF-ANC is shown to have better noise reduction in Fig.~\ref{fig:6 case2} (a) and (b), the power of its output signal is far over the rated output power of 0.8, which in practice causes output saturation and produces an undesirable phenomenon such as divergence and nonlinearity. Furthermore, the proposed algorithm not only suppresses the output power to be within the rated output power but also produces an anti-noise signal that is highly linear with the interference signal as depicted in Fig.~\ref{fig:6 case2} (c), and thus ensures the stability of the system.

\subsection{Real noise with real path}
\begin{figure}[!t]
    \centering
    \includegraphics[width = 8.5cm,height= 7cm]{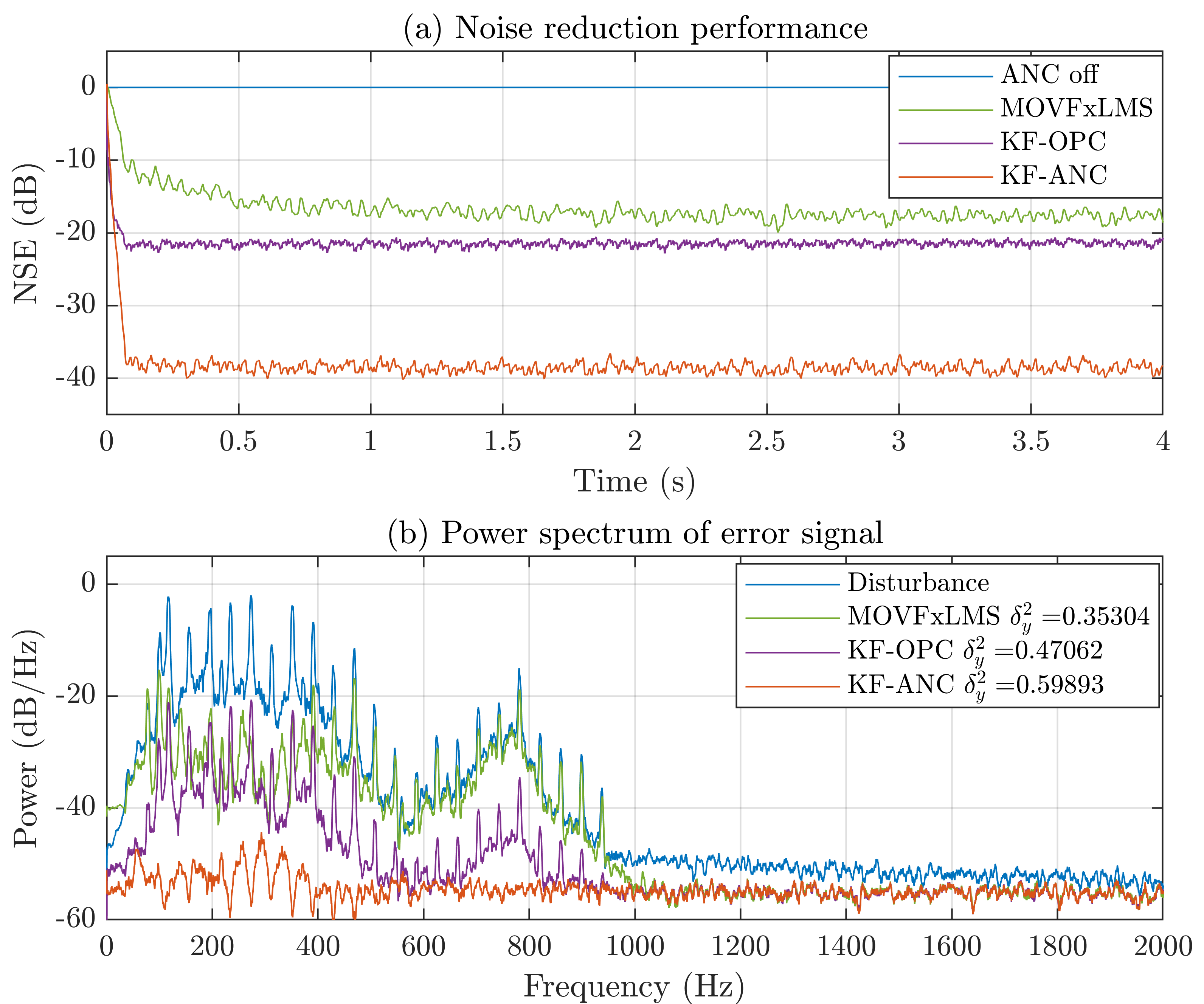}
    \caption{Compressor noise reduction performance under real measured path: (a) Noise reduction performance with different algorithms, and (b) The power spectrum of error signal with different algorithms where the rated power of the system should be $\rho_o = 0.5$.}
    \label{fig:7 case3}
\end{figure}

To manifest the practicality of the KF-OPC algorithm, we use a real primary path and secondary path, measured from the duct, and the recorded compressor noise as primary noise. From the results obtained in Fig.~\ref{fig:7 case3} that the conventional KF-ANC has the best noise reduction performance but disregards the constraint on output power. The KF-OPC algorithm uses a constraint factor to confine the output power to within the rated power, which sacrifices a certain amount of noise reduction but ensures the stability of the system. Compared to the MOVFxLMS algorithm, KF-OPC method has a faster convergence speed. In addition, since the output power of KF-OPC is closer to the rated output power it has better noise reduction than MOVFxLMS, demonstrating superior performance.

\vspace{-0.1cm}
\section{Conclusion}
\label{sec:conclusion}

This paper introduces a Kalman filter with an output power constraint (KF-OPC) method for the ANC system. A constraint factor is utilized to re-scale the disturbance, which will be treated as the measurement for the Kalman filter to conduct the adaption. As a result, the system's output power can be confined to a certain value to avoid output saturation problems. The numerical simulations demonstrate that the proposed algorithm restricts the power of the output signal without affecting the speed of convergence, ensuring stability and linearity.

\vspace{-0.2cm}
\section{Acknowledgement}
This research is supported by the Singapore Ministry of Education, Academic Research Fund Tier 2, under research grant MOE-T2EP20221-0014.

\newpage
\bibliographystyle{IEEEbib}
\bibliography{refs}

\begin{thebibliography}{10}

\bibitem{KuoANC1999}
Sen~M Kuo and Dennis~R Morgan,
\newblock ``Active noise control: a tutorial review,''
\newblock {\em Proceedings of the IEEE}, vol. 87, no. 6, pp. 943--973, 1999.

\bibitem{Elliott1993ANC}
S.~J. Elliott and P.~A. Nelson,
\newblock ``Active noise control,''
\newblock {\em IEEE Signal Processing Magazine}, vol. 10, no. 4, pp. 12--35, 1993.

\bibitem{Kajikawa2012recent}
Yoshinobu Kajikawa, Woon-Seng Gan, and Sen~M. Kuo,
\newblock ``Recent advances on active noise control: open issues and innovative applications,''
\newblock {\em APSIPA Transactions on Signal and Information Processing}, vol. 1, 2012.

\bibitem{lam2021ten}
Bhan Lam, Woon-Seng Gan, DongYuan Shi, Masaharu Nishimura, and Stephen Elliott,
\newblock ``Ten questions concerning active noise control in the built environment,''
\newblock {\em Building and Environment}, vol. 200, pp. 107928, 2021.

\bibitem{shen2023implementations}
Xiaoyi Shen, Dongyuan Shi, Santi Peksi, and Woon-Seng Gan,
\newblock ``Implementations of wireless active noise control in the headrest,''
\newblock in {\em INTER-NOISE and NOISE-CON Congress and Conference Proceedings}. Institute of Noise Control Engineering, 2023, vol. 265, pp. 3445--3455.

\bibitem{hasegawa2017headrest}
Rina Hasegawa, Yoshinobu Kajikawa, Cheng-Yuan Chang, and Sen~M Kuo,
\newblock ``Headrest application of multi-channel feedback active noise control with virtual sensing technique,''
\newblock in {\em INTER-NOISE and NOISE-CON Congress and Conference Proceedings}. Institute of Noise Control Engineering, 2017, vol. 255, pp. 3513--3524.

\bibitem{buck2018performance}
Jan Buck, Sergej Jukkert, and Delf Sachau,
\newblock ``Performance evaluation of an active headrest considering non-stationary broadband disturbances and head movement,''
\newblock {\em The Journal of the Acoustical Society of America}, vol. 143, no. 5, pp. 2571--2579, 2018.

\bibitem{zhang2022robust}
Zeqiang Zhang, Ming Wu, Lan Yin, Chen Gong, Jiajie Wang, Shuang Zhou, and Jun Yang,
\newblock ``Robust feedback controller combined with the remote microphone method for broadband active noise control in headrest,''
\newblock {\em Applied Acoustics}, vol. 195, pp. 108815, 2022.

\bibitem{jung2019local}
Woomin Jung, Stephen~J Elliott, and Jordan Cheer,
\newblock ``Local active control of road noise inside a vehicle,''
\newblock {\em Mechanical Systems and Signal Processing}, vol. 121, pp. 144--157, 2019.

\bibitem{jia2020hybrid}
Zibin Jia, Xu~Zheng, Quan Zhou, Zhiyong Hao, and Yi~Qiu,
\newblock ``A hybrid active noise control system for the attenuation of road noise inside a vehicle cabin,''
\newblock {\em sensors}, vol. 20, no. 24, pp. 7190, 2020.

\bibitem{jiang2021modified}
Yao Jiang, Shuming Chen, Feihong Gu, Hao Meng, and Yuntao Cao,
\newblock ``A modified feedforward hybrid active noise control system for vehicle,''
\newblock {\em Applied Acoustics}, vol. 175, pp. 107816, 2021.

\bibitem{wang2023experimental}
Shuping Wang, Hang Li, Pengju Zhang, Jiancheng Tao, Haishan Zou, and Xiaojun Qiu,
\newblock ``An experimental study on the upper limit frequency of global active noise control in car cabins,''
\newblock {\em Mechanical Systems and Signal Processing}, vol. 201, pp. 110672, 2023.

\bibitem{lee2021review}
Hsiao~Mun Lee, Yuting Hua, Zhaomeng Wang, Kian~Meng Lim, and Heow~Pueh Lee,
\newblock ``A review of the application of active noise control technologies on windows: Challenges and limitations,''
\newblock {\em Applied Acoustics}, vol. 174, pp. 107753, 2021.

\bibitem{huang2011active}
Huahua Huang, Xiaojun Qiu, and Jian Kang,
\newblock ``Active noise attenuation in ventilation windows,''
\newblock {\em The Journal of the Acoustical Society of America}, vol. 130, no. 1, pp. 176--188, 2011.

\bibitem{murao2017mixed}
Tatsuya Murao, Chuang Shi, Woon-Seng Gan, and Masaharu Nishimura,
\newblock ``Mixed-error approach for multi-channel active noise control of open windows,''
\newblock {\em Applied Acoustics}, vol. 127, pp. 305--315, 2017.

\bibitem{shi2019practical}
Dongyuan Shi, Woon-Seng Gan, Jianjun He, and Bhan Lam,
\newblock ``Practical implementation of multichannel filtered-x least mean square algorithm based on the multiple-parallel-branch with folding architecture for large-scale active noise control,''
\newblock {\em IEEE Transactions on Very Large Scale Integration (VLSI) Systems}, vol. 28, no. 4, pp. 940--953, 2019.

\bibitem{Morgan1980FxLMS}
D.~Morgan,
\newblock ``An analysis of multiple correlation cancellation loops with a filter in the auxiliary path,''
\newblock {\em IEEE Transactions on Acoustics, Speech, and Signal Processing}, vol. 28, no. 4, pp. 454--467, 1980.

\bibitem{lee2009subband}
Kong-Aik Lee, Woon-Seng Gan, and Sen~M Kuo,
\newblock {\em Subband adaptive filtering: theory and implementation},
\newblock John Wiley \& Sons, 2009.

\bibitem{luo2023delayless}
Zhengding Luo, Dongyuan Shi, Woon-Seng Gan, and Qirui Huang,
\newblock ``Delayless generative fixed-filter active noise control based on deep learning and bayesian filter,''
\newblock {\em IEEE/ACM Transactions on Audio, Speech, and Language Processing}, 2023.

\bibitem{zhang2023deep}
Hao Zhang and DeLiang Wang,
\newblock ``Deep mcanc: A deep learning approach to multi-channel active noise control,''
\newblock {\em Neural Networks}, vol. 158, pp. 318--327, 2023.

\bibitem{qiu2001study}
Xiaojun Qiu and Colin~H Hansen,
\newblock ``A study of time-domain fxlms algorithms with control output constraint,''
\newblock {\em The Journal of the Acoustical Society of America}, vol. 109, no. 6, pp. 2815--2823, 2001.

\bibitem{guo2024survey}
Yu~Guo, Dongyuan Shi, Xiaoyi Shen, Junwei Ji, and Woon-Seng Gan,
\newblock ``A survey on adaptive active noise control algorithms overcoming the output saturation effect,''
\newblock {\em Signal Processing}, p. 109525, 2024.

\bibitem{shi2019two}
DongYuan Shi, Woon-Seng Gan, Bhan Lam, and Chuang Shi,
\newblock ``Two-gradient direction fxlms: An adaptive active noise control algorithm with output constraint,''
\newblock {\em Mechanical Systems and Signal Processing}, vol. 116, pp. 651--667, 2019.

\bibitem{tobias2005leaky}
Orlando~Jos{\'e} Tobias and Rui Seara,
\newblock ``Leaky-fxlms algorithm: Stochastic analysis for gaussian data and secondary path modeling error,''
\newblock {\em IEEE Transactions on speech and audio processing}, vol. 13, no. 6, pp. 1217--1230, 2005.

\bibitem{shi2021optimal}
Dongyuan Shi, Woon-Seng Gan, Bhan Lam, Shulin Wen, and Xiaoyi Shen,
\newblock ``Optimal output-constrained active noise control based on inverse adaptive modeling leak factor estimate,''
\newblock {\em IEEE/ACM Transactions on Audio, Speech, and Language Processing}, vol. 29, pp. 1256--1269, 2021.

\bibitem{shi2019optimal}
Dongyuan Shi, Bhan Lam, Woon-Seng Gan, and Shulin Wen,
\newblock ``Optimal leak factor selection for the output-constrained leaky filtered-input least mean square algorithm,''
\newblock {\em IEEE Signal Processing Letters}, vol. 26, no. 5, pp. 670--674, 2019.

\bibitem{lai2023mov}
Chung~Kwan Lai, Dongyuan Shi, Bhan Lam, and Woon-Seng Gan,
\newblock ``Mov-modified-fxlms algorithm with variable penalty factor in a practical power output constrained active control system,''
\newblock {\em IEEE Signal Processing Letters}, 2023.

\bibitem{diniz1997adaptive}
Paulo~SR Diniz et~al.,
\newblock {\em Adaptive filtering}, vol.~4,
\newblock Springer, 1997.

\bibitem{lopes1999kalman}
Paulo~AC Lopes and Moises Piedade,
\newblock ``The kalman filter in active noise control,''
\newblock in {\em INTER-NOISE and NOISE-CON Congress and Conference Proceedings}. Institute of Noise Control Engineering, 1999, vol. 1999, pp. 1111--1124.

\bibitem{petersen2008kalman}
Cornelis~D Petersen, Rufus Fraanje, Ben~S Cazzolato, Anthony~C Zander, and Colin~H Hansen,
\newblock ``A kalman filter approach to virtual sensing for active noise control,''
\newblock {\em Mechanical Systems and Signal Processing}, vol. 22, no. 2, pp. 490--508, 2008.

\bibitem{van2013multi}
Sjoerd van Ophem and Arthur~P Berkhoff,
\newblock ``Multi-channel kalman filters for active noise control,''
\newblock {\em The Journal of the Acoustical Society of America}, vol. 133, no. 4, pp. 2105--2115, 2013.

\bibitem{fraanje2005fast}
Rufus Fraanje, Ali~H Sayed, Michel Verhaegen, and Niek~J Doelman,
\newblock ``A fast-array kalman filter solution to active noise control,''
\newblock {\em International Journal of Adaptive Control and Signal Processing}, vol. 19, no. 2-3, pp. 125--152, 2005.

\bibitem{fabry2018active}
Johannes Fabry, Stefan Liebich, Peter Vary, and Peter Jax,
\newblock ``Active noise control with reduced-complexity kalman filter,''
\newblock in {\em 2018 16th International Workshop on Acoustic Signal Enhancement (IWAENC)}. IEEE, 2018, pp. 166--170.

\bibitem{luo2023gfanc}
Zhengding Luo, Dongyuan Shi, Xiaoyi Shen, Junwei Ji, and Woon-Seng Gan,
\newblock ``Gfanc-kalman: Generative fixed-filter active noise control with cnn-kalman filtering,''
\newblock {\em IEEE Signal Processing Letters}, 2023.

\bibitem{yu2024implementation}
Guo Yu,
\newblock ``Implementation of kalman filter approach for active noise control by using matlab: Dynamic noise cancellation,''
\newblock {\em arXiv preprint arXiv:2402.06896}, 2024.

\bibitem{li2023distributed}
Tianyou Li, Siyuan Lian, Sipei Zhao, Jing Lu, and Ian~S Burnett,
\newblock ``Distributed active noise control based on an augmented diffusion fxlms algorithm,''
\newblock {\em IEEE/ACM Transactions on Audio, Speech, and Language Processing}, vol. 31, pp. 1449--1463, 2023.

\end{thebibliography}

\end{document}